\documentclass{IEEEtran4PSCC}
\ifCLASSINFOpdf
   \usepackage[pdftex]{graphicx}
\else
   \usepackage[dvips]{graphicx}
\fi
\usepackage{enumitem}
\usepackage[cmex10]{amsmath}
\usepackage[caption=false]{subfig}
\usepackage{lipsum} 

\usepackage{color}

\makeatletter
\let\old@ps@headings\ps@headings
\let\old@ps@IEEEtitlepagestyle\ps@IEEEtitlepagestyle
\def\psccfooter#1{%
    \def\ps@headings{%
        \old@ps@headings%
        \def\@oddfoot{\strut\hfill#1\hfill\strut}%
        \def\@evenfoot{\strut\hfill#1\hfill\strut}%
    }%
    \def\ps@IEEEtitlepagestyle{%
        \old@ps@IEEEtitlepagestyle%
        \def\@oddfoot{\strut\hfill#1\hfill\strut}%
        \def\@evenfoot{\strut\hfill#1\hfill\strut}%
    }%
    \ps@headings%
}
\makeatother

\psccfooter{%
        \parbox{\textwidth}{\hrulefill \\ \small{21st Power Systems Computation Conference} \hfill \begin{minipage}{0.2\textwidth}\centering \vspace*{4pt} \includegraphics[scale=0.06]{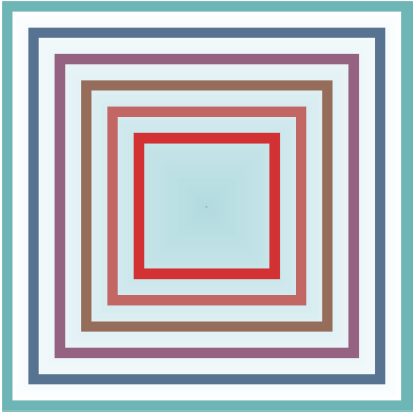}\\\small{PSCC 2020} \end{minipage} \hfill \small{Porto, Portugal --- June 29 -- July 3, 2020}}%
}

\begin{document}
%
\title{Hybrid Transmission Distribution Co-simulation: Frequency Regulation using Battery Energy Storage}

\author{
\IEEEauthorblockN{Gayathri Krishnamoorthy and Anamika Dubey}
\IEEEauthorblockA{School of Electrical Engineering \& Computer Science\\
Washington State University\\
Pullman, USA\\
g.krishnamoorthy@wsu.edu, anamika.dubey@wsu.edu\\}
}


\maketitle

\begin{abstract}
Battery energy storage systems (BESS) are proving to be an effective solution in providing frequency regulation services to the bulk grid. However, there are several concerns for the transmission/distribution system operators (TSO/DSO) with the frequent dispatching of the distribution-connected fast-responding storage systems. Unfortunately, the existing decoupled models for transmission and distribution (T\&D) simulations are unable to capture the complex interactions between the two systems especially concerning frequency regulation problems due to rapidly varying distribution-connected distributed energy resources (DERs). In this paper, an iteratively coupled T\&D hybrid co-simulation framework is developed to facilitate the planning studies for the system operators to help integrate distribution-connected BESS in providing frequency regulation services in response to highly variable DERs such as photovoltaic generation (PVs). Specifically, the proposed framework helps evaluate: (1) the effects of distribution-connected DERs/PVs on the response of system's automatic generation control (AGC) response, and (2) highlights the use of BESS in providing frequency regulation services using integrated T\&D model. The proposed framework is demonstrated using IEEE 9-bus transmission system model (operating in dynamics mode) coupled with multiple EPRI Ckt-24 distribution system models (operating in quasi-static mode). It is shown that the proposed co-simulation framework helps better visualize the system AGC response and frequency regulation especially in the presence of high-levels of DER generation variability requiring frequent dispatch of BESS. 
\end{abstract}



\section{Introduction}
The increasing penetrations of renewable distributed energy resources (DERs) and energy storage systems (ESS) is proving to be a promising solution in the movement towards a decarbonized grid \cite{ref1,ref2}. With the integration of a significant amount of DERs specifically, photovoltaic systems (PVs), the system operators are facing critical challenges with regards to maintaining the supply and demand balance. It is important to account for the intermittency in generation from these resources as they can adversely affect the grid frequency \cite{ref3, ref5}. Traditionally, a majority of frequency regulation capability is provided by specially equipped generators. Recently, following the FERC order 755 \cite{ref9}, all transmission system operators (TSO) and regional transmission organizations (RTO) in the U.S. have implemented pay-for-performance regulation markets to procure automatic generation control (AGC) services from distribution connected battery energy storage systems (BESS) while accounting for the state of charge (SoC) constraint of BESS in their regulation dispatch \cite{ref11}.    

However, there are several concerns for the transmission/distribution system operators (TSO/DSO) with the frequent dispatching of these distribution-connected fast-responding storage systems. First, having uncertainties about the system disturbances and the frequency changes in real-time operation makes it difficult to accurately schedule the BESS \cite{ref7}. Second, the contribution of BESS regulation services in the upcoming time slots depends on their utilization at the current time slot, which the TSO is unaware of \cite{low7}. And third, the AGC scheduling at the TSO level is non-trivial because it depends on the power flow changes of the distribution network, which is affected by charging demand fluctuations in the BESS to provide regulation services \cite{ref8}. To simplify the analysis, currently, the total available distribution-connected BESS (and other non-dispatchable DERs) is aggregated at the transmission and distribution coupling point in planning studies  \cite{ref14, ref15, ref16}. However, the aggregated modeling not only ignores the individual constraints for BESS units but also does not appropriately represent the network constraints at the distribution-level. To successfully enable the participation of dispersed BESS in providing secondary control services specifically, frequency regulation services, an appropriate strategy to distribute AGC dispatch signal among dispersed BESS is required while accounting for their operating constraints. 


Moreover, a high-level of DER penetrations in the distribution system may lead to reverse power flow from distribution to transmission systems that may adversely affect the transmission system operations resulting in frequency regulation problems. Furthermore, with frequent load changes and BESS responding to RegD (fast responding AGC signals) from the TSO, an aggregated battery model at the T\&D coupling point is no longer adequate to represent actual system response. TSOs should be aware of the BESS availability at every connecting point in the distribution system to perform planning studies for AGC response from BESS. This calls for an integrated T\&D planning study for frequency regulation services. While a fully dynamic T\&D model can capture these scenarios, it is unnecessarily complicated \cite{ref5}. This is because, although the transmission system needs to be modeled in a dynamic mode to fully study the bulk grid AGC response, for frequency regulation concerns, the effects of distribution-connected DER generation variability can be captured using quasi-static simulations for distribution systems. Note that a quasi-static T\&D co-simulation platform, that has recently gained significant attention in related literature including our previous work \cite{ref20}, is not sufficient for such studies as it cannot appropriately model grid's frequency response. This calls for a hybrid co-simulation platform that can appropriately model the bulk grid frequency response without unnecessarily complicating the model. It is also important that the tightly coupled co-simulation model \cite{ref20} is used when the BESS are dispersed in the distribution system model as the ACE response obtained from loosely coupled models \cite{HELICS} does not capture the BESS dispatch accurately for system's with high levels of DER penetrations.   

The objective of this paper is to address these and associated challenges in using BESS to provide frequency regulation services for the bulk grid. Specifically, we propose a tightly coupled hybrid T\&D co-simulation framework to analyze integrated T\&D systems due to the BESS dispatch in response to fast regulation AGC signal. To appropriately capture the grid's frequency response, the transmission system is simulated in dynamic mode (in msec) while the distribution system in quasi-static mode (in seconds).  This work adds the following innovations to the existing literature:

\begin{enumerate}[leftmargin=*]
\item \textit{Hybrid T\&D co-simulation platform:} A tightly coupled hybrid T\&D co-simulation platform is developed with transmission system operating in dynamic mode (in msec) to accurately capture the frequency changes due to rapidly varying load demand at the T\&D PCC while the distribution systems is modeled in quasi-static mode (in sec.).

\item \textit{Impacts of PV variability analysis:} The developed hybrid T\&D co-simulation platform is used to understand the PV integration impacts at both transmission and distribution levels. The effects of PV generation variability are studied on the AGC dispatch signals. 

\item \textit{BESS to improve Frequency Regulation:} An approach is detailed to fully exploit the potential of distribution-connected BESS in augmenting transmission system operation by providing frequency regulation. This is achieved by dispatching the ACE signal to BESS for frequency regulation instead of the conventional AGC regulation signal that filters out the fast  load/generation variations. 

\end{enumerate}

The rest of the paper is organized as follows. Section II details the development of the T\&D hybrid co-simulation platform. Section III presents the AGC distribution strategy employed in this work. Section IV details the test system and test cases developed to perform AGC simulations on the tightly coupled hybrid T\&D co-simulation platform with BESS followed by concluding remarks in Section V.
 
\section{Hybrid T\&D co-simulation framework}
The tightly coupled hybrid T\&D co-simulation platform includes four components: the transmission system model, AGC regulation unit at the TSO level, the distribution system model, and the co-simulation interface that co-ordinates the simulation among these individual components. This section details the dynamic transmission system modeling and simulation, quasi-static distribution system modeling and simulation, and the tightly coupled co-simulation interface. The description of AGC regulation unit at the transmission level is explained in Section III. The need for using an integrated T\&D system model to perform AGC simulation studies is also detailed. 

\subsection{Transmission and Distribution System Models}
To manage the frequent load changes, the governor control of the generators at the transmission system responds to balance the supply and demand using the primary frequency control, also called the frequency response, within a few seconds of the supply-demand imbalance. Frequency regulation in power systems corresponds to the secondary frequency control given by AGC which acts in the time scale of minutes after the governor frequency control has responded. To track both the primary and secondary frequency control signals, the transmission system needs to be modeled in the incremental time-step of milliseconds. Thus, in our work, a dynamic model for the transmission system is simulated using power system analysis toolbox (PSAT) in MATLAB.  
\vspace{-0.5cm}
\begin{figure}[ht]
    \centering
    \includegraphics[width=0.5\textwidth]{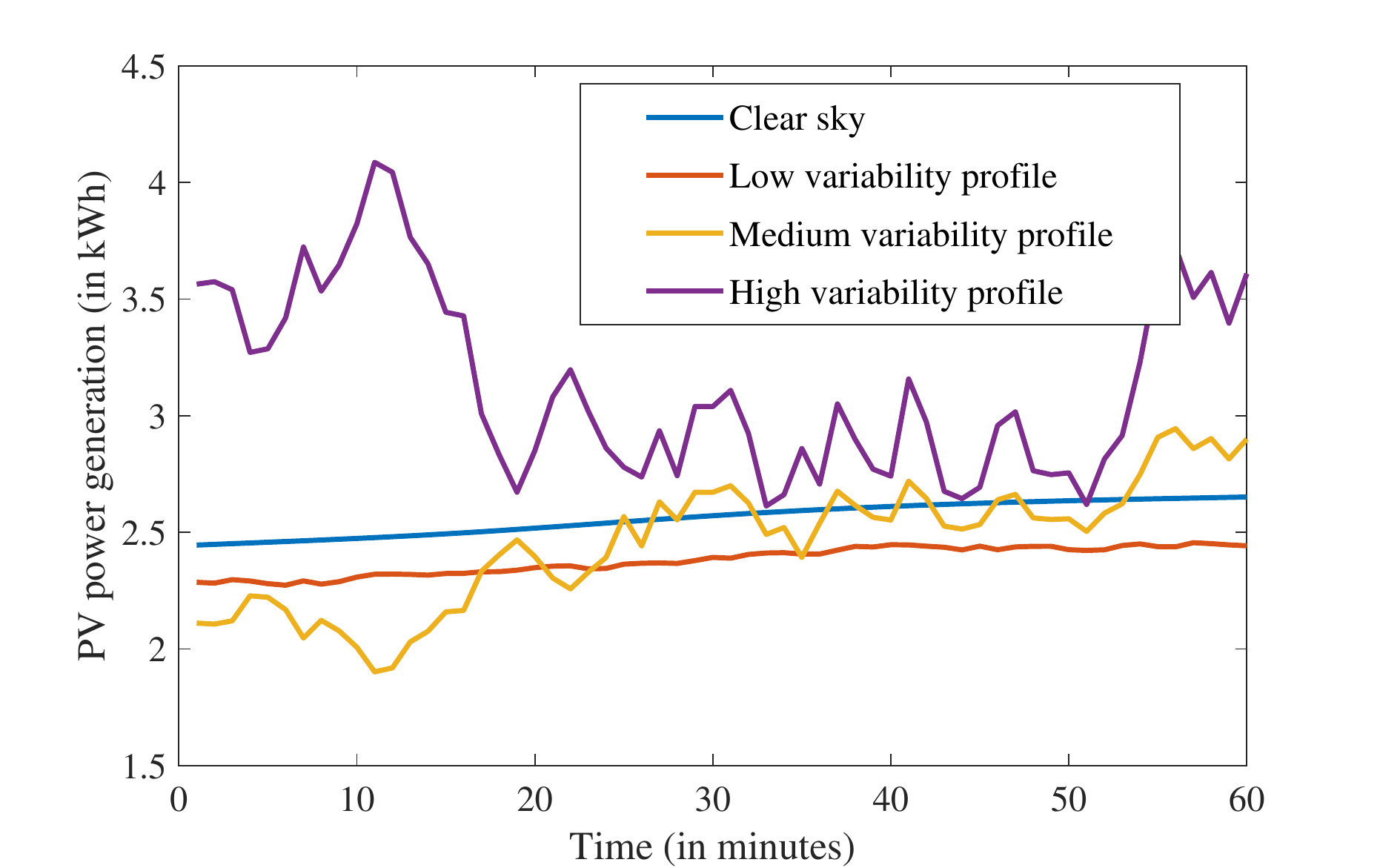}
    \caption{PV generation profile with low, medium, and high variability}
    \label{pvprof}
    \vspace{-5 pt}
\end{figure}

The distribution system is modeled and solved in a three-phase representation using OpenDSS \cite{OpenDSS} to capture the system unbalance conditions. The DER generation scenarios are simulated by modeling distributed PV systems connected to the load buses of the distribution system. Here, the PV deployment cases are unintentionally unbalanced as the PV systems are randomly deployed on the different phases. Please refer to \cite{pvprof} for details on PV deployment. One PV deployment scenario with random size and location of PVs is simulated. The simulated cases for low, medium and high PV variability incur a variability index (VI) of 1.33, 6.29, and 15.58, respectively. \cite{Viref}. A specific 1-hour (12.00-1.00pm) time window of the day is selected for real-time AGC simulations and the corresponding PV profile for the hour is shown in Figure \ref{pvprof}. 

\begin{figure}[ht]
    \centering
    \includegraphics[width=0.5\textwidth]{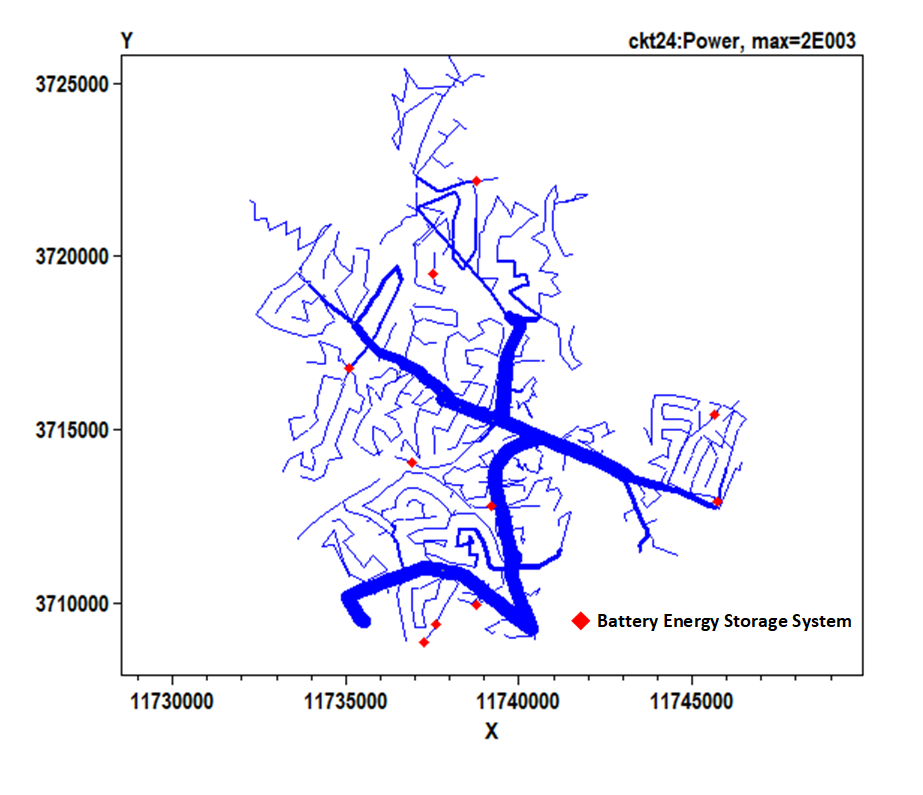}
    \vspace{-1cm}
    \caption{EPRI Ckt-24 distribution feeder with BESS locations}
    \label{ds}
    \vspace{-10 pt}
\end{figure}

For the integrated T\&D test system development, the three load nodes in the IEEE 9-bus transmission system are replaced with EPRI Ckt-24 distribution feeders each. 10 BESS are distributed along each of the connected EPRI Ckt-24 systems as presented in Figure \ref{ds}. Each BESS has a rated power capacity of ±10 kW with 4.21 kWh energy rating. Each connected Ckt-24 feeders can thus replace 300 kW of conventional generation in providing frequency regulation services. The active power output and the SoC of the BESS are monitored through the OpenDSS simulation platform.
\vspace{-0.5cm}

\subsection{Need for an Integrated T\&D System Analysis for AGC Simulation Studies}

The integrated T\&D simulation is required for AGC simulations for multiple reasons. First, with frequent load changes and battery energy storage systems (BESS) responding to fast regulation signals from the TSO, the aggregated battery model and capacity available at the T\&D coupling point is no longer adequate. The TSO should be aware of the BESS availability at every connecting point in the distribution system to appropriately distribute the ACE signal to individual BESS units. Second, a high-level of DER penetrations in the distribution system may lead to reverse power flow from distribution to transmission that may adversely affect the transmission system operations resulting in frequency regulation problems due to power imbalance or other power quality-related issues. Specifically, in the lightly loaded areas, with the increase in DER penetrations, the distributed generation may exceed the local consumption needs, resulting in a reverse power flow condition from individual consumers through the feeders back to the distribution substation and possibly into the transmission system. To assess the associated challenges, a coupled T\&D framework capable of capturing the interactions between the T\&D systems is required. Thus, the analysis done with decoupled T\&D interface may not be accurately reflect grid's operating conditions. This calls for a co-simulation framework that can not only model the impacts of distribution connected resources on bulk grid operations but also the new technologies such as distribution-connected dispersed BESS for their utility in mitigating operational challenges for the bulk grid.

\section{AGC Distribution Strategy}
This section details the development of the AGC regulation unit that simultaneously schedules both BESS and conventional generators to improve grid's frequency regulation; the proposed AGC regulation unit is an integral component of the hybrid T\&D co-simulation framework. The overall AGC signal dispatch for conventional generators and BESS is done by the TSO at the transmission level while the dispatching of fast regulation AGC signal to the distributed BESS in the distribution system is done by the DSO. 

Traditionally, the system AGC regulation is achieved through a proportional-integral (PI) controller with a low-order filter to smooth the fast random variations of the ACE signal which may cause wear and tear on governor motors and turbine valves of conventional AGC units. The low-order filter with a typically large time constant i.e., 1 minute, can reduce the noise at the expense of speed of response which inhibits the inherent benefits of the fast response capability of BESS. Therefore, an independent AGC participation strategy based on ACE signal distribution from the literature is used in this work \cite{ACE_DS}.

\begin{figure}[ht]
    \centering
    \includegraphics[width=0.45\textwidth]{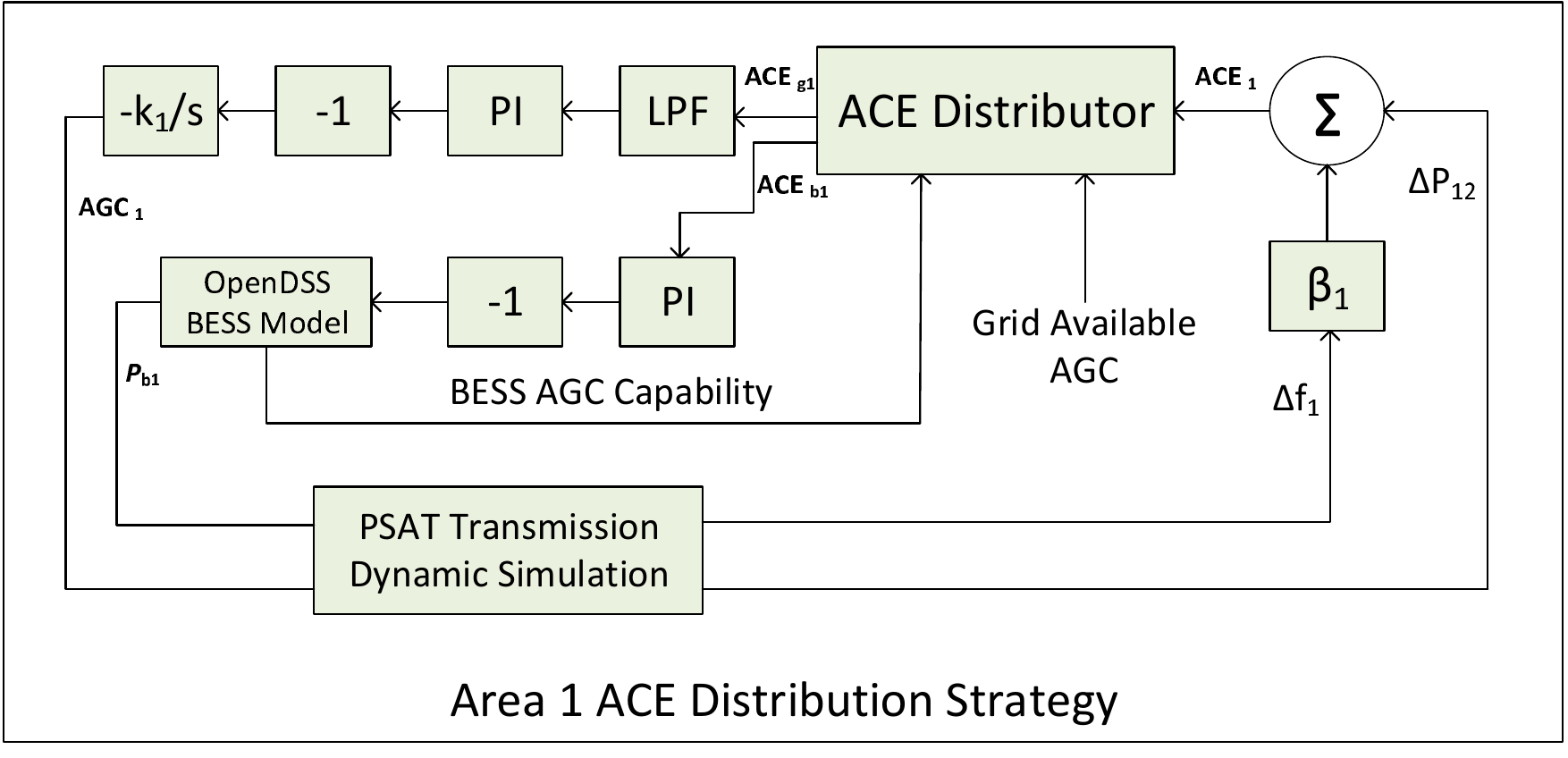}
    \caption{ACE distribution strategy for Area 1}
    \label{ACE_DS}
    \vspace{-5 pt}
\end{figure}

For conventional units, the available AGC capability can be determined purely based on the amount of reserve they possess. However, in the case of BESS, since the amount of stored energy varies over time, the ACE signal is distributed based on the Dynamic Available AGC (DAA) index of the BESS and the assigned ACE signal is sent to the independent AGC PI controller associated with the BESS AGC control loop as shown in Figure \ref{ACE_DS}. Figure \ref{ACE_DS} presents the ACE distribution strategy for the conventional units and the BESS model for one area of the test system. A similar control framework is implemented in other areas of the test system. Here, $\beta_{1}$ represents the droop factor for the area. $\Delta f_{1}$ and $\Delta P_{12}$ represents the frequency deviation and tie-line deviation from the scheduled values in Area 1. The BESS AGC capability represents the DAA index of the BESS connected to Ckt-24 feeders in Area 1. The DAA is 1-minute sustainable charge/discharge capability of BESS considering SOC status and limits. The BESS is continuously calculated and reported at each AGC interval, typically 4-6 seconds. 

To prevent any compromise on the speed of response, the BESS AGC control loop is characterized using a filter with a very small time constant. This allows for fully leveraging the fast response capability of BESS. The figure shows independent control loops for AGC for the conventional generators and BESS for two areas. The ACE index for battery and the other conventional units are calculated as, 

\begin{equation}
    ACE_{bi}=ACE_{i} \times \frac{DAA_{bi}}{DAA_{bi}+AA_{gi}}
\end{equation}
\begin{equation}
    ACE_{gi} = ACE_{i} - ACE_{bi}
\end{equation}

\noindent where $ACE_{bi}$ is the ACE signal for BESS based on availabilities of BESS and conventional units. $DAA_{bi}$ and $AA_{gi}$ are the BESS and conventional units availabilities and $ACE_{gi}$ is the ACE signal for conventional generation units. $ACE_{i}$ is the corresponding area ACE signal. The detailed BESS model available from OpenDSS is used in this study.

\section{Simulation Studies and Results}
Several test cases are simulated to demonstrate the impacts of PV variability on the grid's frequency and AGC response and the proposed secondary frequency control using the proposed ACE distribution strategy for BESS. The simulation study is performed using the developed hybrid T\&D co-simulation platform with low, medium and highly variable PV scenarios. 

\subsection{Test System}
\begin{figure}[ht]
    \centering
    \includegraphics[width=0.45\textwidth]{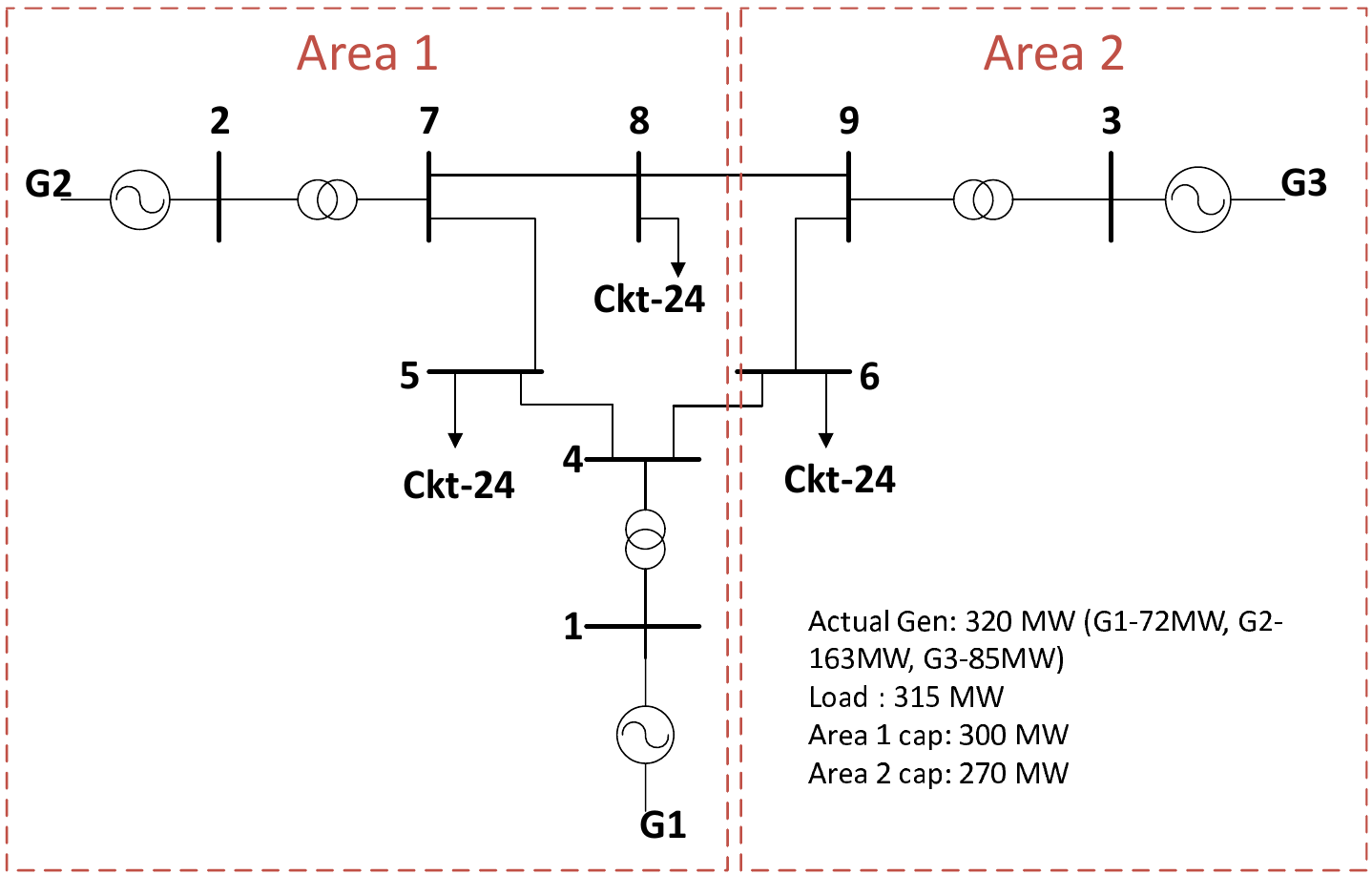}
    \caption{Test System Model}
    \label{9bus}
    \vspace{-10 pt}
\end{figure}

The hybrid T\&D co-simulation test system framework is detailed next. The transmission and distribution systems used in this framework are IEEE 9-bus system \cite{TransTestFeeder} and EPRI Ckt-24 distribution feeder model \cite{DistTestFeeder} available in OpenDSS, respectively. The generator dynamic model for the IEEE 9-bus transmission system is available in PSAT MATLAB toolbox. This system has two areas, three generator buses (including slack bus) and 3 loads as shown in Figure \ref{9bus}. The overall system load is 315 MW with a 3 MW tie-line flow from Area 2 to Area 1. EPRI Ckt-24, a large 6000-bus distribution feeder with 3885 customers and 87\% residential load is used as the distribution system model and is shown in Figure \ref{ds}. The primary side has a voltage level of 13.2 kV and the secondary side voltage level is 480 V and 240 V for 3-phase and 1-phase feeders, respectively. PVs are deployed in EPRI Ckt-24 distribution feeder as described in Section 2A to introduce generation variability. 

For the integrated T\&D test system development, 3 loads in the IEEE 9-bus transmission system are replaced with EPRI Ckt-24 distribution feeders each. 10 BESS are distributed along each of the connected EPRI Ckt-24 systems. Each BESS has a rated power capacity of $\pm$ 10 kW with 4.21 kWh energy rating. Each connected Ckt-24 feeders can thus replace 100 kW of conventional generation in providing frequency regulation services. The active power output and the SoC of the BESS are monitored through the OpenDSS simulation platform. The standard deviation of the system ACE with and without BESS are compared for performance.  


 \subsection{Results and Discussions}
 
 \subsubsection{Transmission and Distribution System Parameters - with and without BESS}
 
\begin{figure}[ht]
    \centering
    \includegraphics[width=0.5\textwidth]{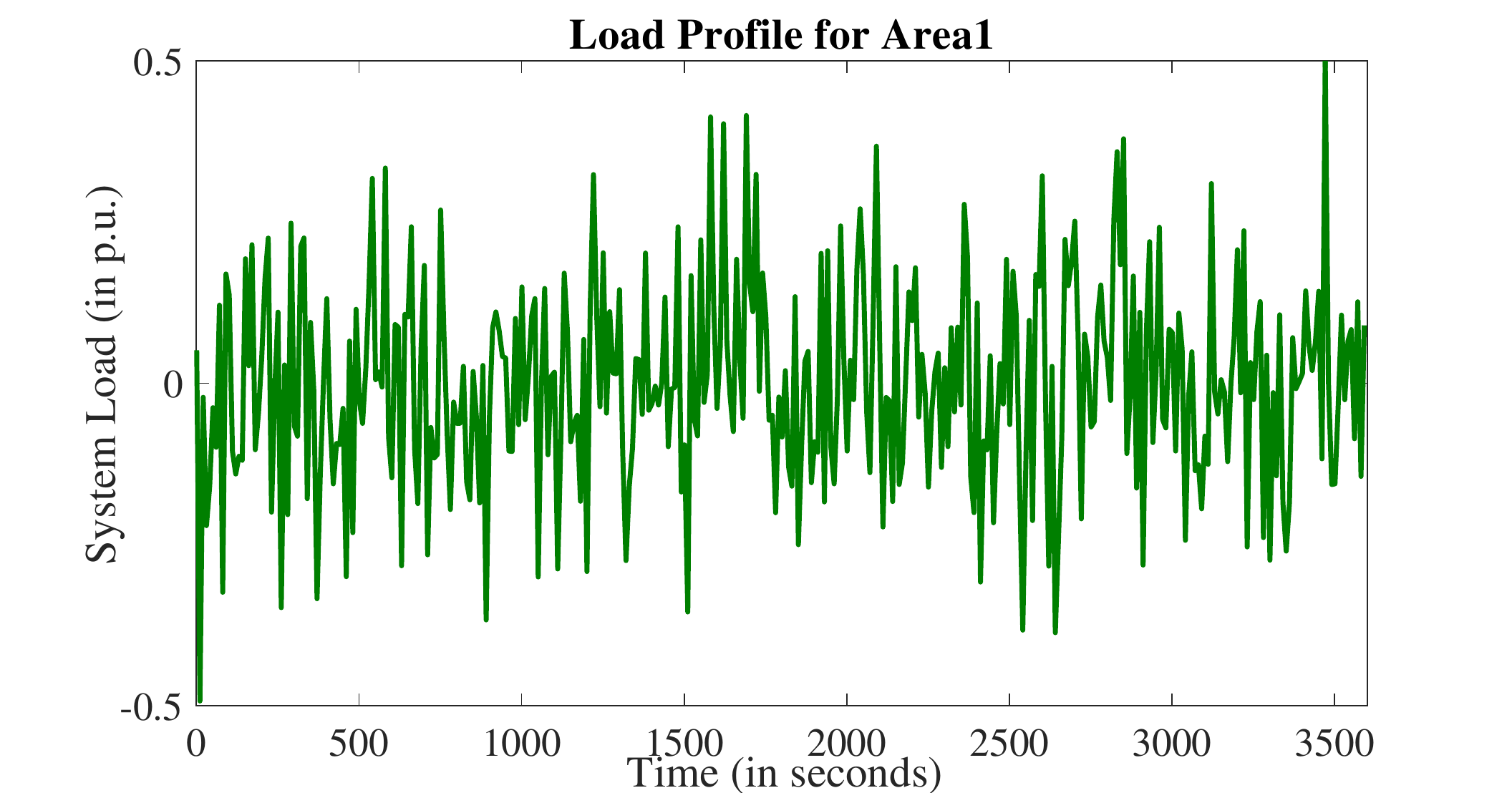}
    \caption{Load profile for Area 1 of the IEEE 9-bus system}
    \label{lp_a1}
    \vspace{-5 pt}
\end{figure}

\begin{figure}[ht]
    \centering
    \includegraphics[width=0.5\textwidth]{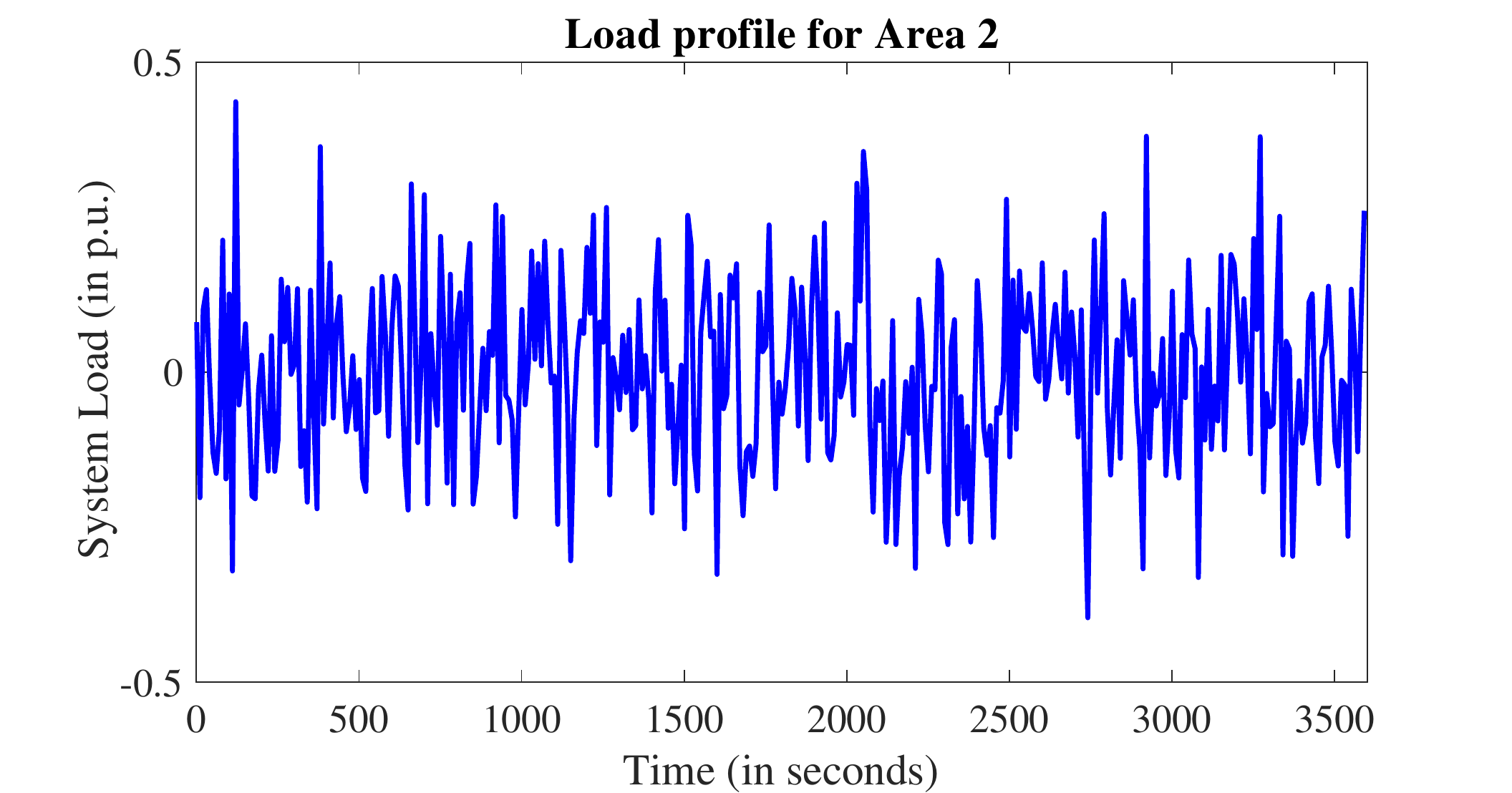}
    \caption{Load profile for Area 2 of the IEEE 9-bus system}
    \label{lp_a2}
    \vspace{-5 pt}
\end{figure}

The following sections utilize BESS in providing frequency regulation services. To demonstrate the benefits of using a hybrid co-simulation platform in replacing the frequency regulation services provided by the conventional generation resources with BESS, the simulation study is performed using ACE distribution strategy as discussed in Section 3. The simulation is carried out for 1-hour of the day (12 pm-1 pm) with high PV variability. The load profiles for the two areas of the test system during the specified hour is presented in Figures \ref{lp_a1} and \ref{lp_a2}. The standard deviation of the system ACE is used for evaluating the performance with and without BESS deployed in the system. 

\paragraph{ Frequency Deviation and ACE Response - with and without BESS}

\begin{figure}[ht]
    \centering
    \includegraphics[width=0.5\textwidth]{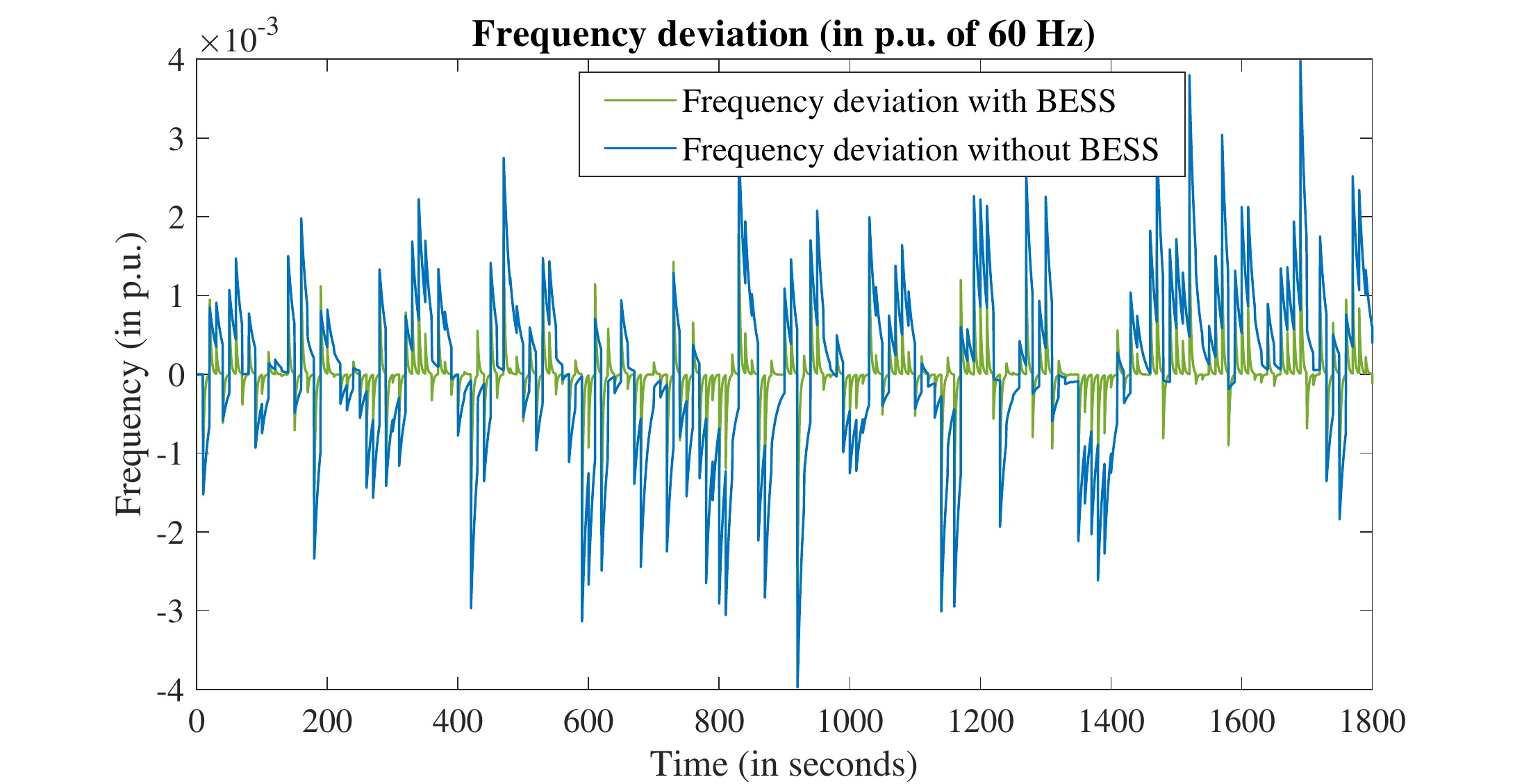}
    \caption{System frequency with and without BESS regulation}
    \label{freq}
    \vspace{-15 pt}
\end{figure}

\begin{figure}[ht]
    \centering
    \includegraphics[width=0.5\textwidth]{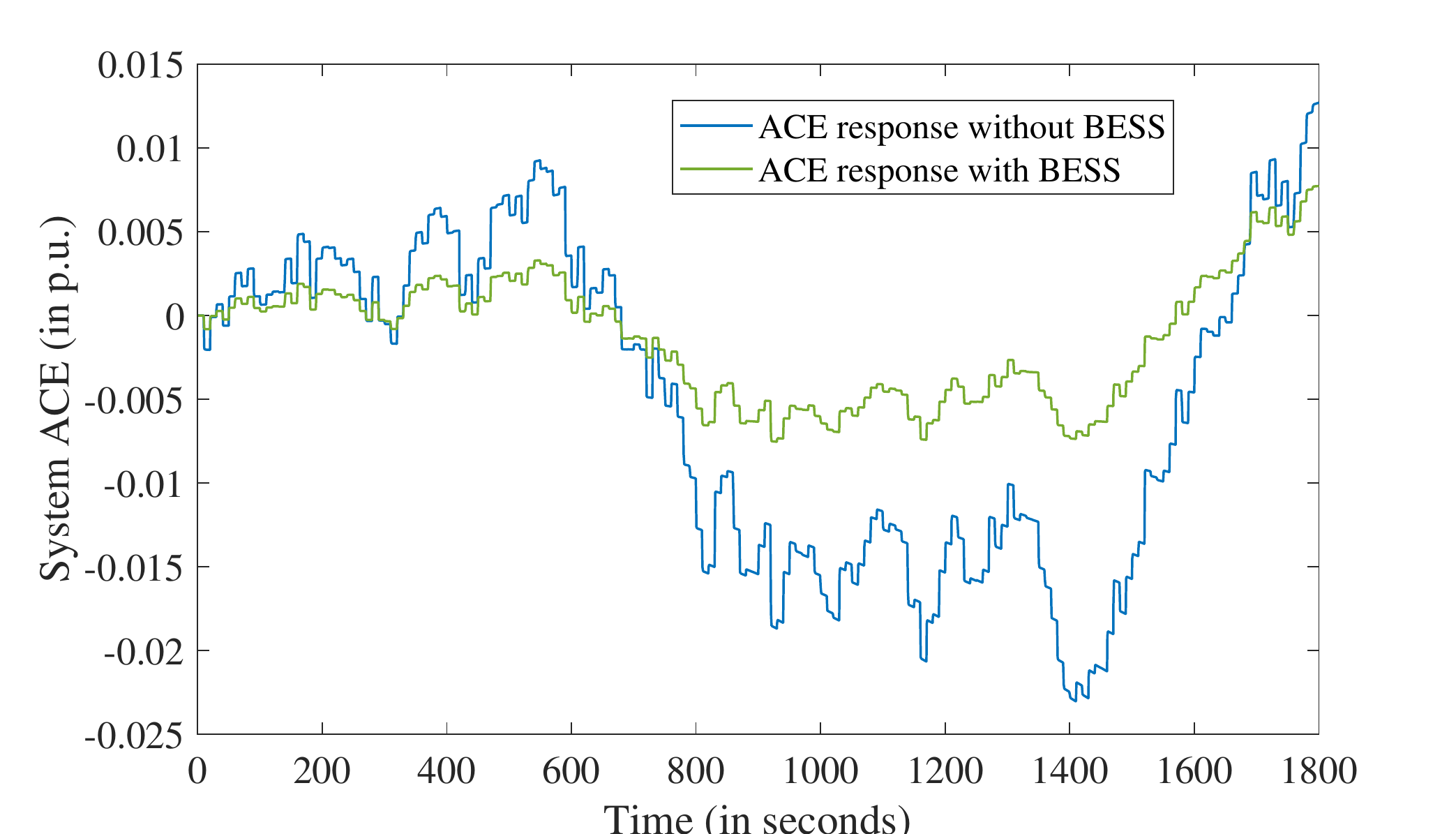}
    \caption{System ACE Response with and without BESS regulation}
    \label{systemACE_case1}
    \vspace{-5 pt}
\end{figure}

The simulation is performed initially without the presence of BESS in the system i.e., the conventional generators in the IEEE 9-bus system (G2, G3) are providing the frequency regulation services. The system ACE response is observed for this initial simulation. Next, BESS is introduced in each of the Ckt-24 feeders connected to the test system and the ACE distribution strategy presented in Section 3 to utilize the BESS services is simulated and the system ACE response is observed again. In addition to the ACE response, the system frequency is also observed in both cases and compared in Figure \ref{freq}. The standard deviation of the system ACE response observed in both simulation cases are compared in Figure \ref{systemACE_case1}.

As seen from the Figures \ref{freq} and \ref{systemACE_case1}, the inclusion of BESS improves both the system frequency regulation and ACE response significantly. It can also be seen that the standard deviation of the system ACE is very low for the simulation case with BESS compared to the one without BESS. This is because the independent AGC control loop for the BESS can better track the variance of the loads in the given test system simulation. In the absence of an integrated T\&D framework with the aggregation of the battery capacity at the PCC not accounting for the SoC constraints, the similar ACE response of the system cannot be observed. This is verified using the simulations performed in the following sections.

\paragraph{ Impacts on Distribution Systems Voltages – with and without BESS}

\begin{figure}[t]
    \centering
    \includegraphics[width=0.5\textwidth]{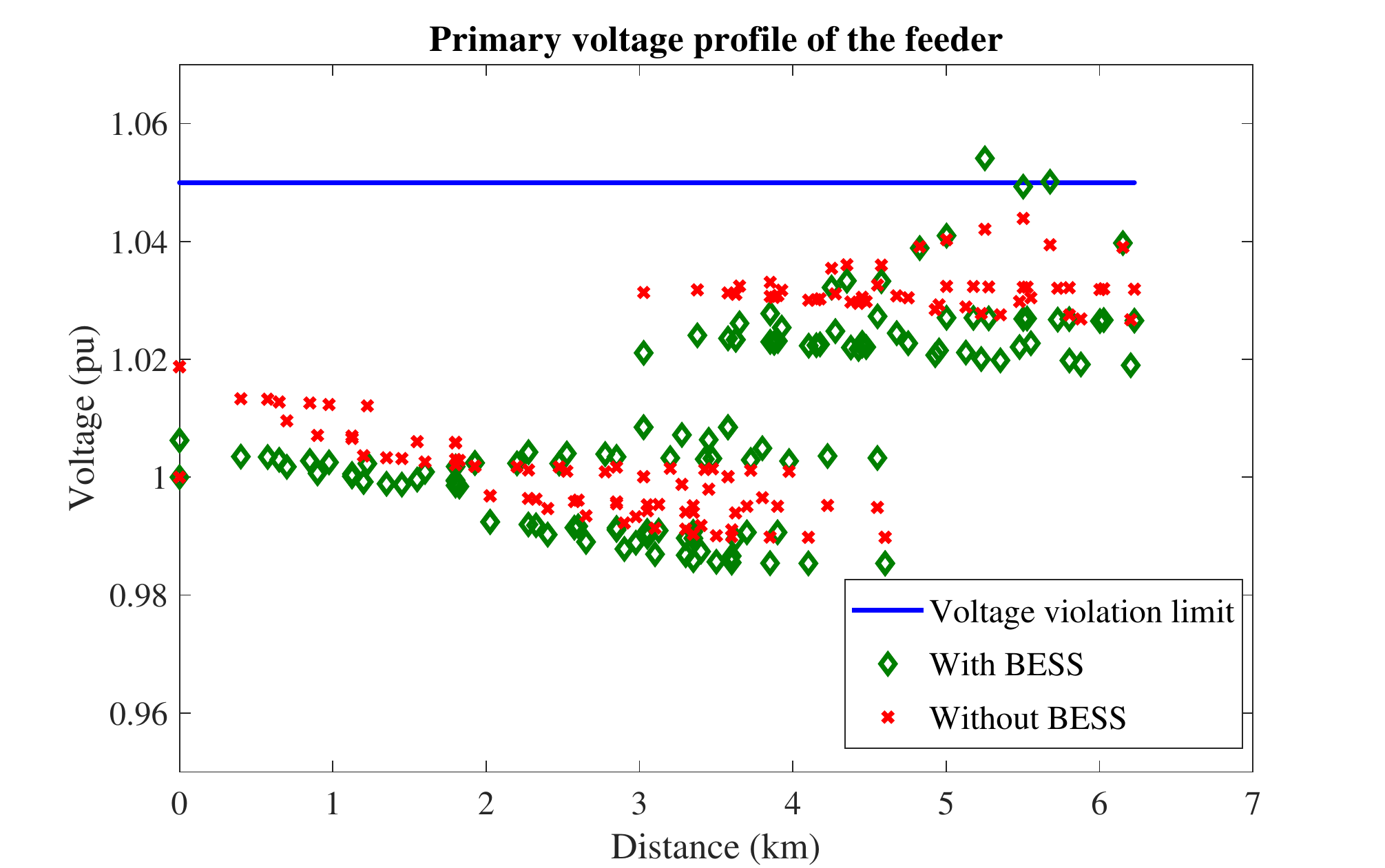}
    \caption{Distribution system voltage profile with and without BESS regulation}
    \label{dsvolt}
    \vspace{-5 pt}
\end{figure}

Next, we observe the distribution system voltages for the simulation case described in the previous section.  The high variability profile for PVs is simulated. The comparison of primary voltage profiles along the distance of the feeder with and without BESS is presented in Figure \ref{dsvolt}. 

As can be observed from the figure, the voltage profiles for the two cases are different. The results are shown for t=480s where the BESS dispatch has led to the violation of the system voltage constraints. This distribution system profile is not observable to TSO when the AGC regulation is conducted by the TSO without complete details of the distribution system model. In this scenario, the hybrid T\&D co-simulation plays an important role for DSO to observe the system voltages when the TSO dispatches distribution-connected BESS that may violate system operating constraints.

\begin{figure}[t]
    \centering
    \includegraphics[width=0.5\textwidth]{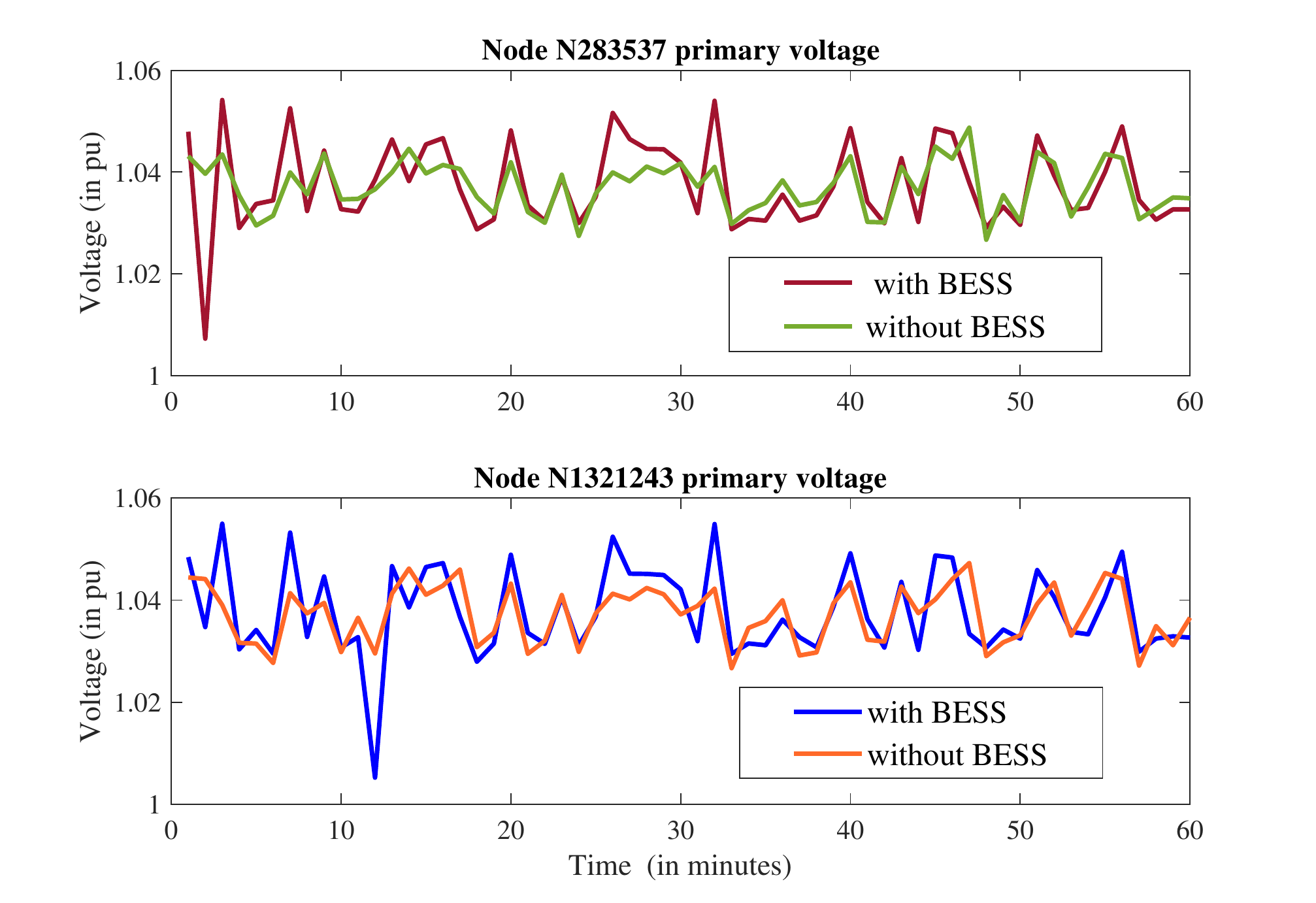}
    \caption{Time-series voltage profile of DS nodes with and without BESS regulation}
    \label{tsvolt}
    \vspace{-15 pt}
\end{figure}

In addition to this, the time-series voltage plots of nodes violating the voltage limit during that simulation hour is presented in Figure \ref{tsvolt}. As can be seen from the figure, with BESS dispatch there is a violation in allowed voltage limits, which cannot be observed by the operator without simulating a complete distribution system model.

\vspace{5pt}
\subsubsection{Aggregated vs. T\&D Co-simulation Models - Impacts on ACE and Frequency Response}

This section further details the need for hybrid T\&D co-simulation platform in performing frequency regulation studies. As discussed previously, the AGC scheduling at the transmission system operator (TSO) level is non-trivial because it depends on the power flow changes of the distribution network, which is affected by the charging demand fluctuations to provide regulation services.  This section demonstrates this issue at the TSO level by simulating the ACE response of the system without a complete distribution system model and compares the same with ACE response for T\&D co-simulation model, where the TSO has detailed information about the distribution system model. 

\begin{figure}[t]
\centering

\subfloat[Low PV variability]{%
  \includegraphics[clip,width=1\columnwidth]{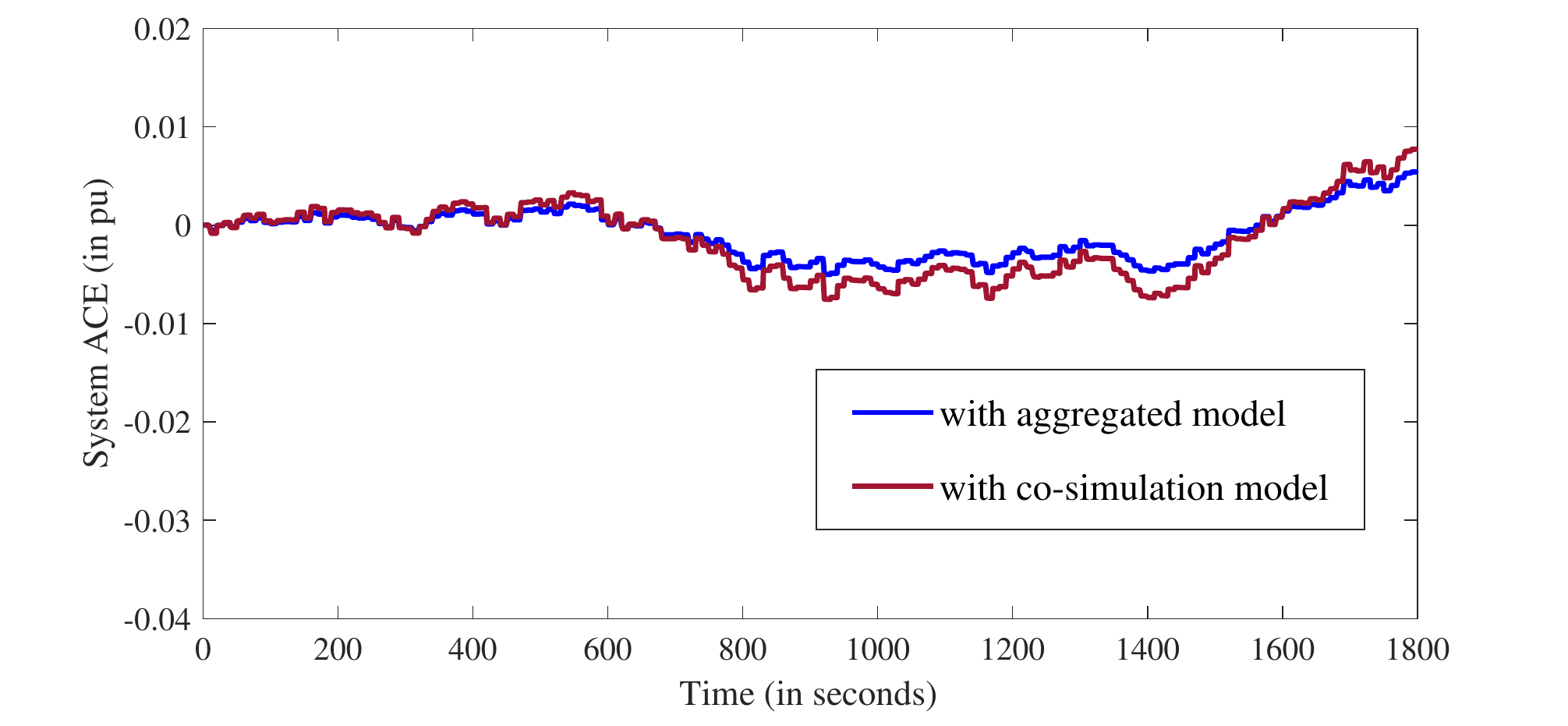}%
}

\subfloat[Medium PV variability]{%
  \includegraphics[clip,width=1\columnwidth]{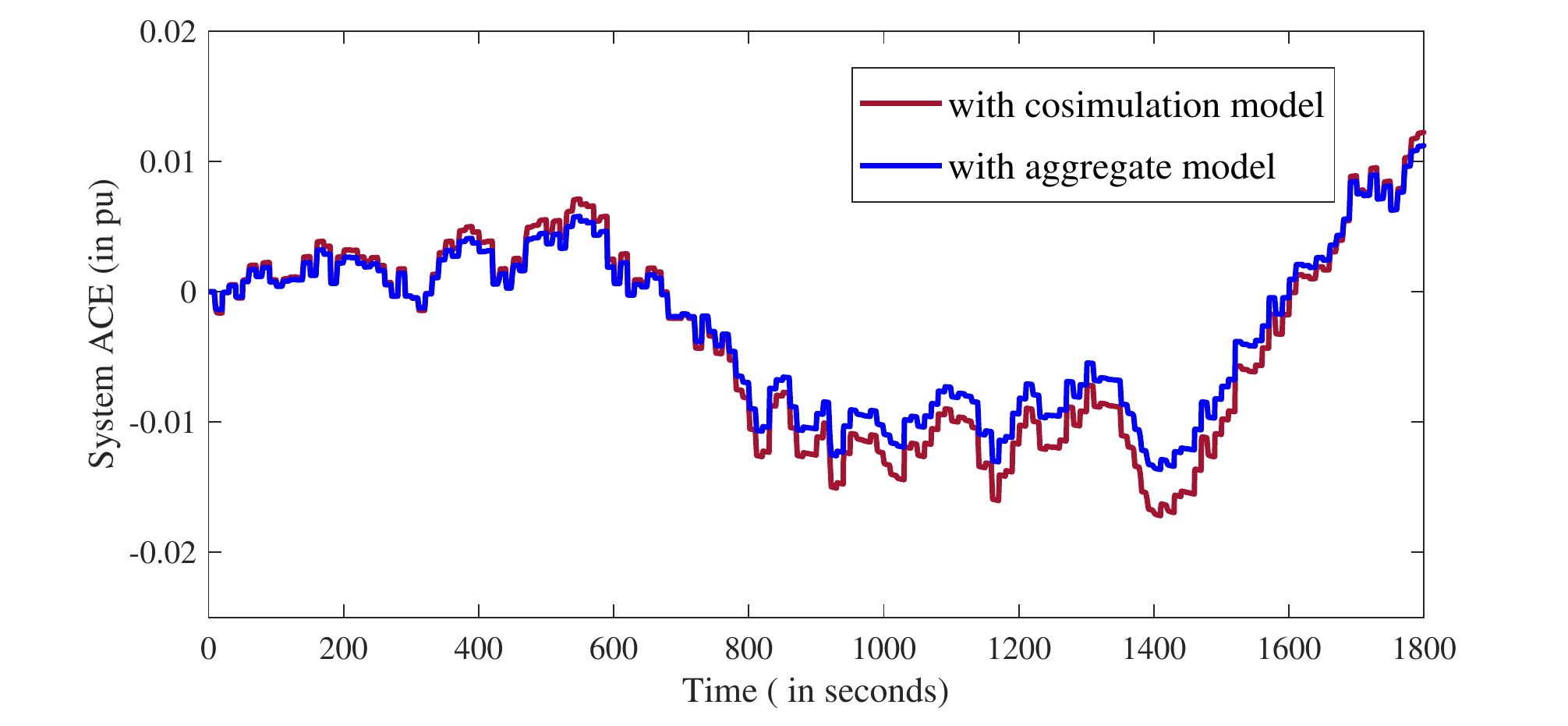}%
}

\subfloat[High PV variability]{%
  \includegraphics[clip,width=1\columnwidth]{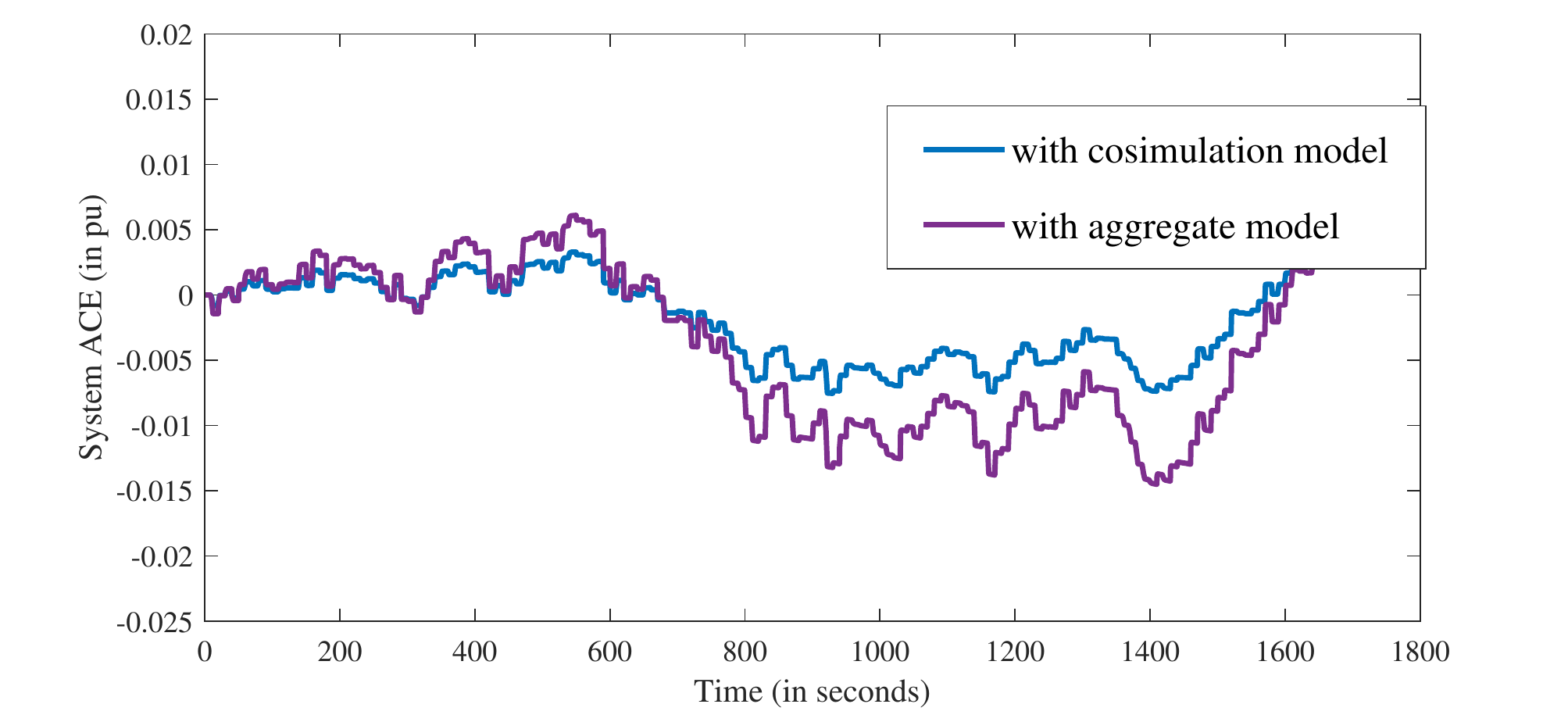}%
}

\caption{ACE response of the system with and without co-simulation platform}
\vspace{2pt}
\label{comp_case2}
\vspace{-15pt}
\end{figure}

\paragraph{Impacts of PV Variability on System ACE and Frequency Deviation – Aggregated vs. T\&D Co-simulation Model}
We simulate two test cases: (1) Aggregated model - the distribution-connected PV systems and loads are aggregated at the transmission load point (PCC), and (2) Co-simulation model - the complete distribution system model is simulated that is coupled with the transmission system model. The simulation is carried out for 1-hour of the day (12 pm-1 pm) with low, medium, and high PV variabilities as discussed in Section 2A. The simulated case for low, medium and high PV variability incurs a variability index (VI) of 1.33, 6.29, and 15.58 [52], and the results are presented in Figure \ref{comp_case2}.

As seen from the figure, with the increase in PV variability, the ACE deviation is increasing in all three cases. Similarly, with the aggregate model where the DS loads are approximated at the transmission load point, the ACE deviation is found to be increasing in all three cases. As mentioned previously, these simulation results demonstrate the need for a complete distribution system model simulation in studying ACE response for system frequency regulation with high DER penetrations. 

\paragraph{BESS to improve System ACE and Frequency Deviation – Aggregated vs. T\&D Co-simulation Model (High Variability Case)}

\begin{figure}[ht]
    \centering
    \includegraphics[width=0.5\textwidth]{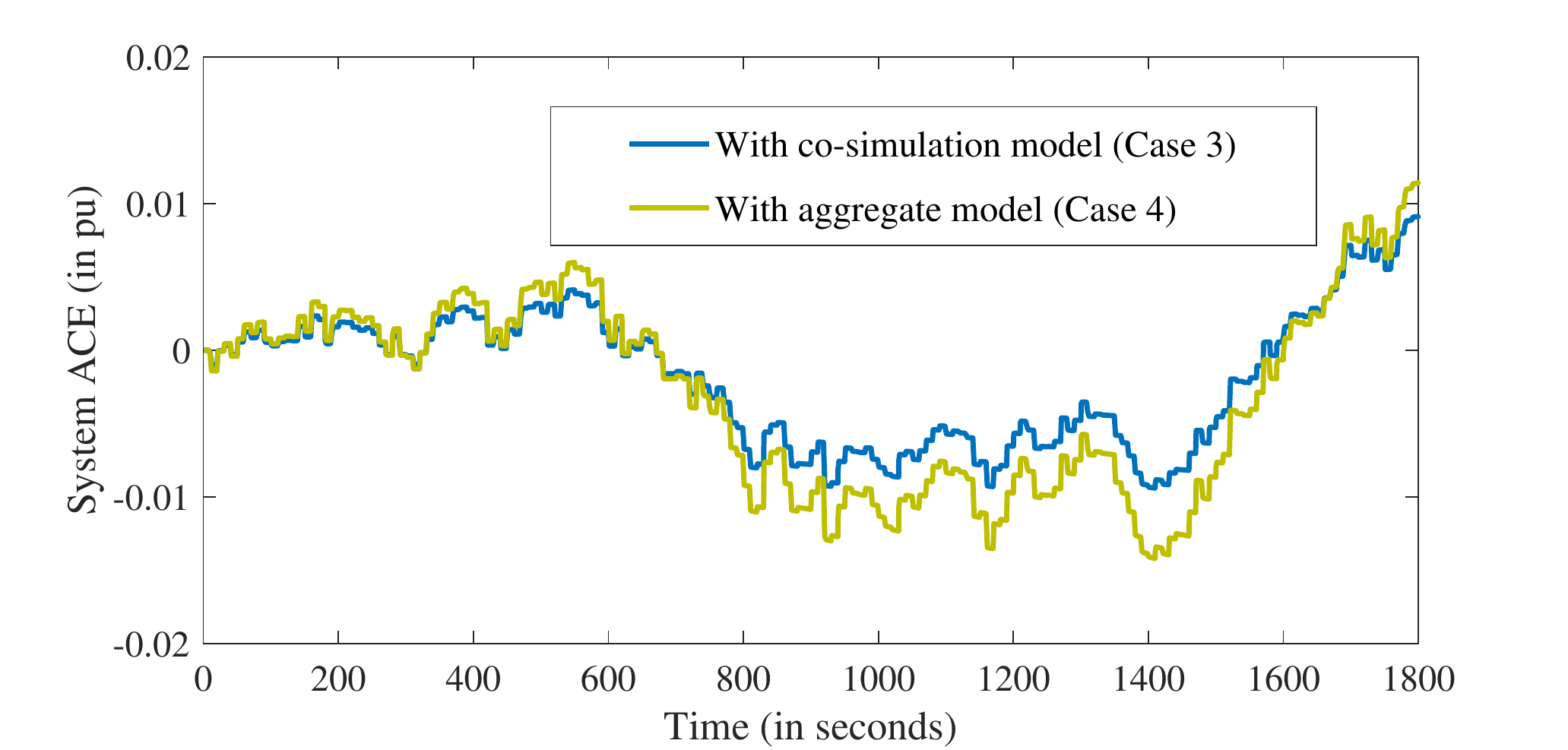}
    \caption{System ACE response with and with co-simulation}
    \label{comp_case3}
    \vspace{-5 pt}
\end{figure}

\begin{figure}[ht]
    \centering
    \includegraphics[width=0.5\textwidth]{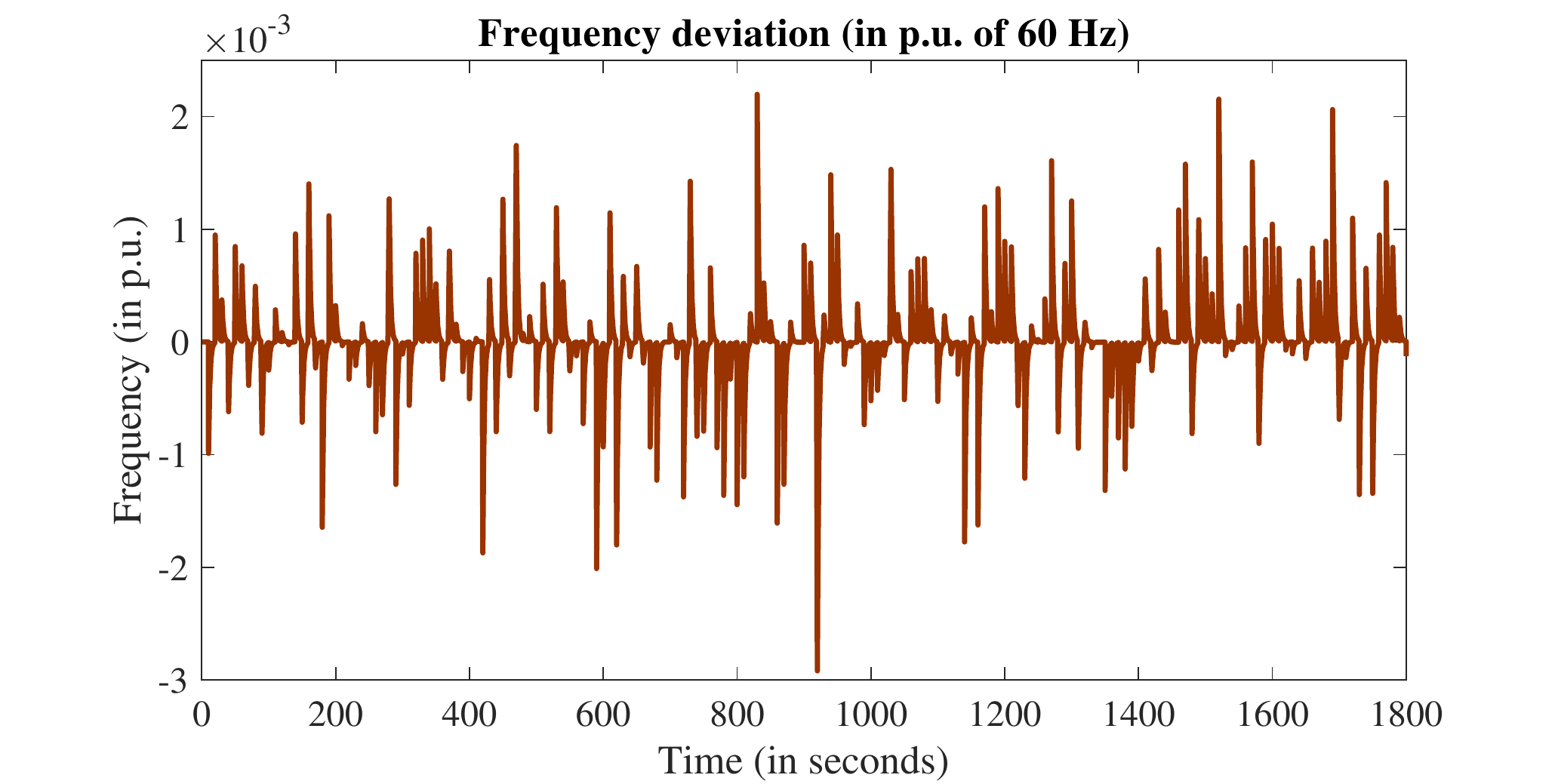}
    \caption{System frequency response with distributed battery using co-simulation (Case 3)}
    \label{freq_resp_case3}
    \vspace{-5 pt}
\end{figure}

In this section, two comparisons are made. One, where the 10 BESS distributed across the distribution system have equal capacity and SoC constraints while in the other, the overall BESS capacity is distributed to 10 batteries differently. Four different cases are simulated. In case 1 (co-simulation model) and case 2 (aggregated model), the BESS are of equal capacity. In case 3 (co-simulation model) and case 4 (aggregated model), the BESS capacities are different. For both aggregated models, the ACE schedule is obtained from the availability of aggregated PV and BESS (i.e., without a complete distribution system model). The resulting schedule is dispatched to individual BESS units in the detailed distribution system model to simulate actual operating conditions. The ACE response of the system for cases 3 and 4 is presented in Figure \ref{comp_case3} and the frequency response for case 3 is presented in Figure \ref{freq_resp_case3}.

As can be seen from Figure \ref{comp_case3}, the ACE magnitude and frequency deviation in the co-simulation model is better than the aggregated model. This is because the schedule of the resources obtained with the aggregated model can be inaccurate while dispatching to a set of batteries with different capacities.  It can be observed from Figure \ref{freq_resp_case3} that the system frequency response in AGC regulation with distributed BESS and hybrid T\&D co-simulation model can better model the AGC requirements of achieving near-zero frequency deviation.

\vspace{-0.5cm}

\subsubsection{ACE Response with LC vs. TC Interaction Models}
\begin{figure}[t]
    \centering
    \includegraphics[width=0.5\textwidth]{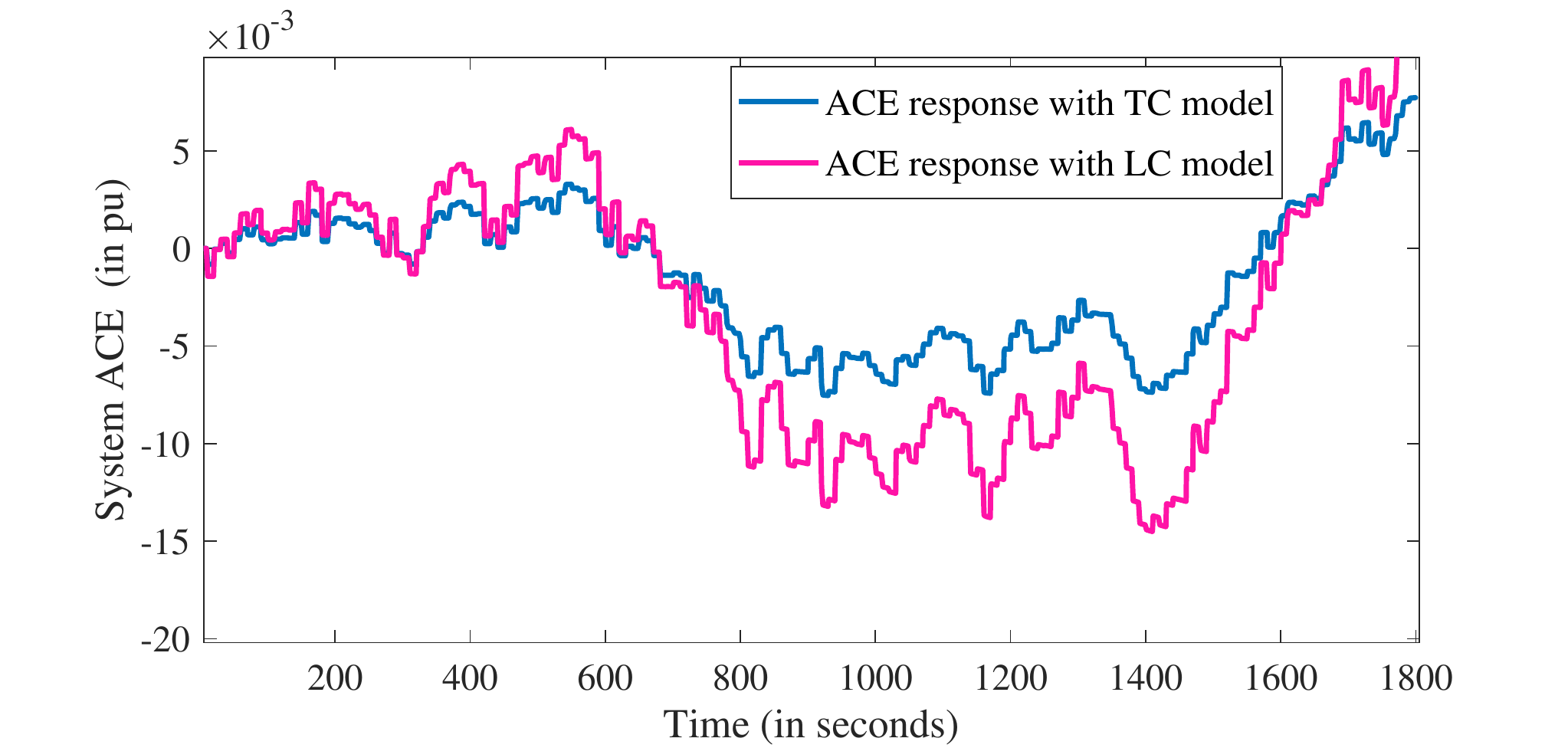}
    \caption{System ACE response from LC vs. TC integration models}
    \label{comp_case4}
    \vspace{-10 pt}
\end{figure}

This simulation is carried out to evaluate the impact of T\&D coupling strength on the accuracy of frequency regulation simulations. Here, the same simulation conditions presented in section 1 are utilized to compare loosely coupled and tightly coupled T\&D co-simulation models. In the LC model, the interchange at the PCC happens only once \cite{HELICS}, while in the TC model the interface variables converge before moving to the next time step \cite{ref20}. The comparison results are presented in Figure \ref{comp_case4}. From the observation, it is clear that the ACE response is better (less standard deviation) in the TC case study. This is because the TC model captures the interactions between transmission and distribution systems accurately even in cases with high PV variabilities.

\section{Conclusion}

This work aims at developing an integrated T\&D platform to leverage the benefits of distribution-connected BESS in providing frequency regulation services for the bulk grid. A hybrid T\&D co-simulation platform is proposed where the transmission system is simulated in dynamic mode in 1-msec interval while the distribution system is simulated in a quasi-static mode in 1-sec interval. The BESS is distributed among multiple locations in the distribution feeder. The frequency regulation operation also termed as secondary frequency control, performed using AGC is operated every 4 seconds. The simulations are carried out for cases with low, medium, and high PV variability, and the results are compared with and without the presence of BESS, with aggregated and distributed BESS modeling and the frequency regulation and system ACE responses are observed. It is shown that the hybrid T\&D co-simulation platform helps better capture the potential of the distribution-connected BESS in providing the frequency regulation responses. This work will be further extended to study and mitigate the problems faced by the distribution systems operators such as violations of distribution-level operational constraints due to the dispatch of fast responding BESS.

\bibliographystyle{ieeetr}
\bibliography{references}



\end{document}